\documentclass{article}%
\usepackage{amsmath}
\usepackage{graphicx}%
\usepackage{amsfonts}%
\usepackage{amssymb}
\begin{document}

\author{D.L. Kovrizhin, L.A. Maksimov\\Moscow, Kurchatov sq.1 RRC Kurchatov Institute}
\title{``Cherenkov radiation'' of a sound in a Bose-condensed gas}

\begin{abstract}
In terms of linearized Gross-Pitaevskii equation we have studied the process
of sound emission arises from a supersonic particle motion in a Bose-condensed
gas. By analogy with the method used for description of Vavilov-Cherenkov
phenomenon, we have found a friction work created by the particle generated
condensate polarization. For comparison we have found radiation intensity of
excitations. Both methods gives the same result.

\end{abstract}
\date{23 February 2001}
\maketitle

\section{Introduction}

The recent experiments \cite{Bradley} of Bose-Einstein condensation (BEC) in
trapped alkali-metal atomic gases have stimulated a great interest on the
problem \cite{pitaevskii}. One of the reasons of such an interest is the fact
that nonlinear Shr\"{o}dinger equation (Gross-Pitaevskii equation) is well
suited for describing the fundamental characteristics of an interacting
Bose-gas at zero temperature and also appears in other physical problems:
theory of superconductivity, theory of elementary particles, nonlinear optics
\cite{Talanov} etc.

Examination of \ fast particle's energy loss is one of the popular
experimental technique in condensed matter physics. Such experiments also can
give an additional information about properties of an ultracold Bose-gas.

It is known, when a particle moves in a matter with a velocity$\ v$ above the
supersonic speed $c$ which equals to the speed of excitations, the matter
begins to radiate. In electrodynamics this phenomenon names Vavilov-Cherenkov
radiation \cite{landau8}, in aerodynamics such an effect appears from a
supersonic motion of a body, in condensed matter physics --- this is a polaron
problem \cite{Feinman}.

In the paper we use linearized Gross-Pitaevsky equation to study the process
of radiation of the excitations in a Bose-condensate, arises from a supersonic
particle motion. Under this consideration, the Bose-condensate excitations are
described as the excitations of a classic nonrelativistic scalar field, same
as radiation of a classic electromagnetic field in a continuous matter. As it
was shown \cite{Stringari}, the GPE can be reduced to the system of
hydrodynamic equations. But this approach adequately describes only the
long-wave part of the spectrum, which corresponds to the low energy levels of
elementary excitations $\omega\ll\mu.$ As it will be shown below, the maximum
of the radiation is in the region $\omega_{\max}=2\mu(v/c)\sqrt{(v/c)^{2}-1}.$
So that, direct investigation of the GPE gives more detailed description of
the problem. Note that to find an energy loss of a particle it is enough to
solve the problem by the methods of quantum mechanics using ``Fermi's golden
rule.'' The result will be the same.

\section{Particle driven condensate polarization}

The state of the condensate (with a fixed chemical potential $\mu)$, is
described by the macroscopic wave-function $\Psi\left(  \mathbf{r},t\right)
,$ which satisfies the GPE equation \cite{landau9} ($\hbar=1$),
\[
i\frac{\partial}{\partial t}\Psi\left(  \mathbf{r},t\right)  =\left(
\frac{\hat{p}^{2}}{2m}+V\left(  \mathbf{r},t\right)  -\mu\right)  \Psi\left(
\mathbf{r},t\right)  +\int d\mathbf{r}_{1}U\left(  \mathbf{r}-\mathbf{r}%
_{1}\right)  \Psi^{\ast}\left(  \mathbf{r}_{1},t\right)  \Psi\left(
\mathbf{r}_{1},t\right)  \Psi\left(  \mathbf{r},t\right)  .
\]
The interaction energy of Bose-gas atoms $U\left(  \mathbf{r}-\mathbf{r}%
_{1}\right)  =U_{0}\delta\left(  \mathbf{r}-\mathbf{r}_{1}\right)  $ and
potential energy of its interaction with the straightly and uniformly moving
particle $V\left(  \mathbf{r},t\right)  =V_{0}e^{st}\delta\left(
\mathbf{r}-\mathbf{v}t\right)  ,s=+0$. In the last expression we use an
adiabatic switching of interaction to exclude automatically advanced solutions
. Wave-function of the ground state of a static gas without a particle
corresponding to zero energy (counted from the chemical potential) doesn't
depend on the coordinates and equals $\Psi_{0}=\sqrt{\mu/U_{0}}.$ In the
presence of a particle condensate becomes heterogeneous --- ``polarizes.'' The
``polarization'', in general, depends on time $\psi\left(  \mathbf{r}%
,t\right)  =\Psi\left(  \mathbf{r},t\right)  -\Psi_{0},$ in linear
approximation it satisfies inhomogeneous equation:%
\begin{align}
i\frac{\partial}{\partial t}\psi\left(  \mathbf{r},t\right)   &  =\left(
\frac{\hat{p}^{2}}{2m}+\mu\right)  \psi\left(  \mathbf{r},t\right)  +\mu
\psi^{\ast}\left(  \mathbf{r},t\right)  +be^{st}\delta\left(  \mathbf{r}%
-\mathbf{v}t\right)  ,\label{l_gross}\\
b  &  =V_{0}\Psi_{0},\;\;\left[  b\right]  =[r^{3/2}/t],s=+0.\nonumber
\end{align}
Let's solve the equation (\ref{l_gross}) using Fourier's method. We present
the potential and the $\delta-$function as a plane waves decomposition,%
\begin{gather}
\psi\left(  \mathbf{r},t\right)  =\int\frac{d^{3}k}{\left(  2\pi\right)  ^{3}%
}e^{i\mathbf{kr-}i\omega t+st}\psi\left(  \mathbf{k}\right)  ,\label{f_psi}\\
\delta\left(  \mathbf{r}-\mathbf{v}t\right)  e^{st}=\int\frac{d^{3}\mathbf{k}%
}{\left(  2\pi\right)  ^{3}}e^{i\mathbf{kr-}i\omega t+st},\;\;\omega
=\mathbf{kv.} \label{f_delta}%
\end{gather}
Substituting (\ref{f_psi}) and (\ref{f_delta}) into equation (\ref{l_gross}),
we get%
\begin{equation}
(\omega+is-\xi)\psi\left(  \mathbf{k}\right)  =\mu\psi^{\ast}\left(
-\mathbf{k},\right)  +b,\;\;\xi=\frac{k^{2}}{2m}+\mu.
\end{equation}
Combining this equation with complex conjugated one, we find%

\begin{gather}
\psi\left(  \mathbf{k}\right)  =b\frac{S\left(  \mathbf{k}\right)  }%
{(\omega+is)^{2}-\varepsilon^{2}\left(  k\right)  },\label{psi_1}\\
S\left(  \mathbf{k}\right)  =\frac{k^{2}}{2m}+\omega,\quad\varepsilon
^{2}\left(  k\right)  =\xi^{2}-\mu^{2}=c^{2}k^{2}+\left(  \frac{1}{2m}%
k^{2}\right)  ^{2},\quad c^{2}=\frac{\mu}{m}. \label{111}%
\end{gather}
The poles of (\ref{psi_1}) describes the Bogolyubov's oscillations spectrum
$\varepsilon\left(  k\right)  $ of the Bose-condensate. Substitution of
(\ref{psi_1}) into (\ref{f_psi}) gives an important expression for the field
of condensate ``polarization'',%
\begin{equation}
\psi\left(  \mathbf{r},t\right)  =b\int\frac{d^{3}k}{\left(  2\pi\right)
^{3}}e^{i\mathbf{kr-}i\omega t+st}\frac{S\left(  \mathbf{k}\right)  }%
{(\omega+is)^{2}-\varepsilon^{2}\left(  k\right)  },\;\;\omega=\mathbf{kv}.
\label{psi_2}%
\end{equation}
With a slow particle motion the field decays exponentially at a large distance
from the particle. The poles of (\ref{psi_1}) can lead to radiation,
\[
\mathbf{kv=\pm}\varepsilon\left(  k\right)
\]
Thus, for appearance of \ radiation in a sound spectrum region ($\varepsilon
\left(  k\right)  =ck$) we have the condition,
\[
kv\cos\theta_{0}=ck\;\;\cos\theta_{0}=\frac{c}{v}.
\]
It gives the equation of the front of sound radiation. With a particle motion
parallel to $z$ axis, we have%
\begin{equation}
z=vt-a\sin\theta_{0},\;\;r=a\cos\theta_{0},\;\;0\leq a<\infty. \label{front}%
\end{equation}
If we place a center of \ spherical coordinates at the point of the particle
then there is no field at right of the front in the region $0<\theta
<\theta_{0}+\frac{1}{2}\pi.$ Properly speaking if we consider the dispersion
of sound speed then high-frequency field in the region is not zero, but only
the low-frequency part of the spectrum%
\[
c\left(  k\right)  =\frac{\varepsilon\left(  k\right)  }{k}=\sqrt{c^{2}%
+\frac{1}{4m^{2}}k^{2}}<v
\]
gives contribution to radiation. The dispersion of the sound phase velocity
$c\left(  k\right)  $ plays the same role as the dispersion of \ inductivity,
cutting the radiation frequency on top,
\begin{gather}
c\left(  k_{\max}\right)  =v,\\
\omega_{\max}=k_{\max}v=\varepsilon\left(  k_{\max}\right)  =2mv\sqrt
{v^{2}-c^{2}}%
\end{gather}

\section{Fast particle energy loss}

In first order of perturbation theory, the potential energy of the particle in
the field of condensate ``polarization'' is%
\begin{align*}
\delta E  &  =\int d\mathbf{r}\Psi^{\ast}\left(  \mathbf{r},t\right)  V\left(
\mathbf{r}-\mathbf{R}\right)  \Psi\left(  \mathbf{r},t\right)  ,\\
&  =b\Psi_{0}+2b\operatorname{Re}\psi\left(  \mathbf{R}\right)  ,
\end{align*}
where $\mathbf{R}(t)$ --- is the radius vector of the moving particle. The
energy produces the drag force which acts on the particle from its created
field%
\[
\mathbf{F}=-\frac{d}{d\mathbf{R}}\delta E=-2b\operatorname{Re}\left[
\frac{d\psi}{d\mathbf{R}}\right]  _{\mathbf{r}=\mathbf{v}t}.
\]
This results in particle energy loss%

\[
\dot{E}=\mathbf{Fv}=-2b\mathbf{v}\operatorname{Re}\left[  \frac{d\psi
}{d\mathbf{R}}\right]  _{\mathbf{r}=\mathbf{v}t},
\]
that retires to infinity as a form of condensate excitations. Using the
expression (\ref{psi_2}) we find the radiation intensity%
\[
I=-\dot{E}=2b^{2}\operatorname{Re}\int\frac{d^{3}k}{\left(  2\pi\right)  ^{3}%
}\left(  i\mathbf{kv}\right)  \frac{\dfrac{k^{2}}{2m}+\omega}{(\omega
+is)^{2}-\varepsilon^{2}\left(  k\right)  }.
\]
To integrate the expression let's use Sokhotsky formula%
\[
\frac{1}{(\omega+is)^{2}-\varepsilon^{2}}=P\frac{1}{\omega^{2}-\varepsilon
^{2}}-\frac{i\pi}{2\omega}[\delta\left(  \omega-\varepsilon\right)
+\delta\left(  \omega+\varepsilon\right)  ]
\]
The principal value of the integral gives no contribution into energy loss.
Thus,%
\begin{equation}
I=\frac{b^{2}}{4\pi}\int\limits_{0}^{\infty}k^{2}dk\int\limits_{-1}^{1}%
d\cos\theta\left(  \dfrac{k^{2}}{2m}+\omega\right)  [\delta\left(
\omega-\varepsilon\right)  +\delta\left(  \omega+\varepsilon\right)  ].
\label{11111}%
\end{equation}
Transforming (\ref{11111}) using a new variable $\omega=kv\cos\theta$, we get
the very simple formula,%
\begin{equation}
I=\frac{b^{2}}{4\pi mv}\int\limits_{0}^{k_{\max}}k^{3}dk. \label{009}%
\end{equation}
With the help of the formula which is inversed to (\ref{111}),%
\[
k^{2}=2m(-\mu+\sqrt{\mu^{2}+\varepsilon^{2}}),
\]
we have an expression for the intensity in form of a spectrum decomposition%
\begin{equation}
I=\frac{b^{2}m}{2\pi v}\int\limits_{0}^{\omega_{\max}}\varepsilon\left(
1-\frac{\mu}{\sqrt{\mu^{2}+\varepsilon^{2}}}\right)  d\varepsilon. \label{117}%
\end{equation}

\section{Radiation intensity}

Now, it is interesting to directly evaluate the radiation intensity,
integrating the energy flow $\mathbf{Q}$ by all directions \cite{Zubarev},%

\[
\mathbf{Q}=\operatorname{Re}\left[  \left(  \frac{\mathbf{\hat{P}}}{m}%
\psi\right)  ^{\ast}i\frac{\partial}{\partial t}\psi\right]
\]
where $\mathbf{\hat{P}}=-i\mathbf{\nabla}$ --- is the operator of momentum,
$i\dfrac{\partial}{\partial t}$\ --- the operator of energy. Whole energy flow
through the surface of the cylinder with radius $R\rightarrow\infty$ around
the particle trajectory is%
\begin{equation}
I=\int\mathbf{Q}d\mathbf{S}=2\pi R\operatorname{Re}\int_{-\infty}^{\infty
}dz\left(  \frac{1}{im}\frac{\partial}{\partial R}\psi\right)  ^{\ast}%
i\frac{\partial}{\partial t}\psi\label{flow}%
\end{equation}
To evaluate the integral let's find form of the field (\ref{psi_2}) in the
wave zone in cylindrical coordinates%

\begin{equation}
\psi\left(  \vec{r},t\right)  =\frac{b}{\left(  2\pi\right)  ^{3}}\int
d^{2}k_{\perp}\int\limits_{-\infty}^{\infty}dk_{z}e^{i\vec{k}_{\perp}\vec{R}%
}e^{ik_{z}z}e^{-i\omega t+st}\frac{\frac{1}{2m}(k_{\perp}^{2}+k_{z}%
^{2})+\omega}{(\omega+is)^{2}-\varepsilon^{2}\left(  k\right)  }%
,\;\;\omega=k_{z}v
\end{equation}
Introducing the ``retarded'' time $t^{\prime}=t-\dfrac{z}{v}$, and expressing
the polar angle integral (by variable $\varphi)$ using the definition of
Bessel function%
\[
\int_{0}^{2\pi}e^{ix\cos\varphi}d\varphi=2\pi J_{0}\left(  x\right)  ,
\]
we find%
\[
\psi\left(  \vec{r},t\right)  =\frac{2mb}{\left(  2\pi\right)  ^{2}v}%
\int\limits_{0}^{\infty}k_{\perp}dk_{\perp}J_{0}\left(  k_{\perp}R\right)
\int\limits_{-\infty}^{\infty}d\omega e^{-i\left(  \omega+is\right)
t^{\prime}}\frac{(k_{\perp}^{2}+\frac{\omega^{2}}{v^{2}})+2m\omega}%
{4m^{2}(\omega+is)^{2}-4m\mu(k_{\perp}^{2}+\frac{\omega^{2}}{v^{2}})-\left(
k_{\perp}^{2}+\frac{\omega^{2}}{v^{2}}\right)  ^{2}},
\]
to get more compact expression let's factorize the denominator
\begin{equation}
\psi\left(  \vec{r},t\right)  =-\frac{mb}{\left(  2\pi\right)  ^{2}v}%
\int\limits_{-\infty}^{\infty}d\omega e^{-i\left(  \omega+is\right)
t^{\prime}}F,\;\;\;F=\int\limits_{0}^{\infty}2k_{\perp}dk_{\perp}J_{0}\left(
k_{\perp}R\right)  \frac{k_{\perp}^{2}+\frac{\omega^{2}}{v^{2}}+2m\omega
}{\left(  k_{\perp}^{2}-k_{1}^{2}\right)  \left(  k_{\perp}^{2}+k_{2}%
^{2}\right)  -is\omega} \label{200}%
\end{equation}
where
\begin{align*}
k_{1}^{2}  &  =-\left(  2\mu m+\frac{\omega^{2}}{v^{2}}\right)  +2m\sqrt
{\omega^{2}+\mu^{2}}\\
k_{2}^{2}  &  =\left(  2\mu m+\frac{\omega^{2}}{v^{2}}\right)  +2m\sqrt
{\omega^{2}+\mu^{2}}%
\end{align*}

With small frequencies the positive value of $k_{1}^{2}\sim\omega^{2}$ have a
maximum at $\omega=\sqrt{\left(  mv^{2}\right)  ^{2}-\mu^{2}}$ and becomes
zero at $\omega_{\max}=2mv\sqrt{v^{2}-c}.$ The integral $F\left(
\omega\right)  $ can be represented in a following form%
\begin{equation}
F\left(  \omega\right)  =\int\limits_{0}^{\infty}dk_{\perp}J_{0}\left(
k_{\perp}R\right)  \frac{k_{\perp}^{2}+\frac{\omega^{2}}{v^{2}}+2m\omega
}{4m\sqrt{\omega^{2}+\mu^{2}}}\left[  \frac{2k_{\perp}}{k_{\perp}^{2}-\left(
k_{1}+i\omega s\right)  ^{2}}-\frac{2k_{\perp}}{k_{\perp}^{2}+k_{2}^{2}%
}\right]  \label{112}%
\end{equation}
The last term in the square brackets has imaginary poles, it gives the
exponentially small contribution to the integral at large distances from the
particle trajectory ($R\rightarrow\infty$) and can be neglected. The
expression (\ref{112}) can be transformed to integral by whole real axis
\cite{Ivanenko}.

Let's examine the integral along the closed contour%
\begin{equation}
J=\dfrac{1}{2}\int\limits_{C}f\left(  z^{2}\right)  H_{0}^{\left(  1\right)
}\left(  z\right)  zdz \label{f1}%
\end{equation}
where $H_{0}^{\left(  1\right)  }\left(  zR\right)  $ --- is Hankel function,
analytical in an upper half-plane of complex variable $z$ which have the
following asymptotic behavior on the infinity%
\begin{equation}
H_{0}^{\left(  1\right)  }\left(  z\right)  \approx\sqrt{\frac{2}{\pi z}%
}e^{iz-i\pi/4}. \label{844}%
\end{equation}
The path $C$ consists of the real axis and the infinite semicircle in an
upper-halfplane. The integral along the last contour is exponentially small.
So we can write%
\begin{equation}
J=J_{-}+J_{+},\;\;J_{-}=\dfrac{1}{2}\int\limits_{-\infty}^{0}dzH_{0}^{\left(
1\right)  }\left(  zR\right)  zf\left(  z^{2}\right)  ,\;\;J_{+}=\dfrac{1}%
{2}\int\limits_{0}^{\infty}dzH_{0}^{\left(  1\right)  }\left(  zR\right)
zf\left(  z^{2}\right)  \label{845}%
\end{equation}
Let's transform first integral to a new variable $z=\rho e^{i\varphi}$%

\begin{equation}
J_{-}=-\dfrac{1}{2}\int\limits_{0}^{\infty}d\rho H_{0}^{\left(  1\right)
}\left(  e^{i\pi}\rho R\right)  \rho f\left(  \rho^{2}\right)  . \label{846}%
\end{equation}
Taking into account (\ref{845}) we have for sum (\ref{846})%
\[
J=\dfrac{1}{2}\int\limits_{0}^{\infty}d\rho\rho f\left(  \rho^{2}\right)
[\sqrt{\frac{2}{\pi\rho}}e^{i\rho R-i\pi/4}+\sqrt{\frac{2}{\pi\rho}}e^{-i\rho
R+i\pi/4}]=\int\limits_{0}^{\infty}d\rho\rho f\left(  \rho^{2}\right)
J_{0}\left(  \rho R\right)  .
\]
On the other hand the integral on the closed contour (\ref{f1}) equals the sum
of residues of the poles $z_{n}$ in upper-halfplane%

\begin{equation}
J=2\pi i\sum_{res}\dfrac{1}{2}H_{0}^{\left(  1\right)  }\left(  z_{n}R\right)
z_{n}f\left(  z_{n}^{2}\right)  \label{847}%
\end{equation}
Applying (\ref{847}) to integral (\ref{112}) and remembering that integrand
have one pole at $\left|  \omega\right|  <\omega_{\max}$ close to the real
axis in the upper-halfplane equal to $k_{\perp}=\left(  k_{1}%
\operatorname{sign}\omega+is\right)  $ we have%

\begin{align}
F\left(  \omega\right)   &  =2\pi iH_{0}^{\left(  1\right)  }\left(  p\right)
\frac{k_{1}^{2}+\frac{\omega^{2}}{v^{2}}+2m\omega}{4m\sqrt{\omega^{2}+\mu^{2}%
}}=\pi iH_{0}^{\left(  1\right)  }\left(  p\right)  D\left(  \omega\right)
,\label{115}\\
D\left(  \omega\right)   &  =\frac{k_{1}^{2}+\frac{\omega^{2}}{v^{2}}%
+2m\omega}{2m\sqrt{\omega^{2}+\mu^{2}}}=1+\frac{\omega-\mu}{\sqrt{\omega
^{2}+\mu^{2}}},\;\;p=k_{1}R\operatorname{sign}\omega+is. \label{810}%
\end{align}
Substituting (\ref{115}) into (\ref{200}) and taking into account that at
$\left|  \omega\right|  >\omega_{\max}$ parameter $k_{1}$\ becomes imaginary
and Hankel function is exponentially small, we find $\psi\left(
\mathbf{r},t\right)  $ in the wave zone as a form of single integral by
frequency%
\begin{equation}
\psi\left(  \mathbf{r},t\right)  =-\frac{imb}{8\pi v}\int\limits_{-\omega
_{\max}}^{\omega_{\max}}d\omega e^{-i\left(  \omega+is\right)  t^{\prime}%
}H_{0}^{\left(  1\right)  }\left(  p\right)  D\left(  \omega\right)  .
\label{811}%
\end{equation}
Let's use this expression to find (\ref{flow}). Using (\ref{811}),
(\ref{844}),\ (\ref{810}) we have%
\begin{equation}
-\frac{i}{m}\frac{\partial}{\partial R}\psi=-\frac{imb}{8\pi v}\int
\limits_{-\omega_{\max}}^{\omega_{\max}}d\omega\left(  \frac{k_{1}}%
{m}sign\omega\right)  e^{-i\left(  \omega+is\right)  t^{\prime}}H_{0}^{\left(
1\right)  }\left(  p\right)  D\left(  \omega\right)  \label{p_psi}%
\end{equation}
and%
\begin{equation}
i\frac{\partial\psi}{\partial t}=-\frac{imb}{8\pi v}\int\limits_{-\omega
_{\max}}^{\omega_{\max}}d\omega\left(  \omega\right)  e^{-i\left(
\omega+is\right)  t^{\prime}}H_{0}^{\left(  1\right)  }\left(  p\right)
D\left(  \omega\right)  \label{h_psi}%
\end{equation}
Substituting this expressions into (\ref{flow}) and integrating by $z$ with a
formula%
\[
\int_{-\infty}^{\infty}e^{-i\left(  \omega-\omega^{\prime}\right)  \left(
t-z/v\right)  }dz=2\pi v\delta\left(  \omega-\omega^{\prime}\right)
\]
and using explicit form (\ref{844}) we have%
\begin{align*}
I  &  =4\pi^{2}Rv\left(  \frac{mb}{8\pi v}\right)  ^{2}\int\limits_{-\omega
_{\max}}^{\omega_{\max}}d\omega\frac{\left|  \omega\right|  }{m}k_{1}\frac
{2}{\pi\left|  p\right|  }D\left(  \omega\right)  ^{2}\\
&  =\frac{mb}{8\pi v}^{2}\int\limits_{-\omega_{\max}}^{\omega_{\max}}%
d\omega\left|  \omega\right|  D\left(  \omega\right)  ^{2}=\frac{mb}{2\pi
v}^{2}\int\limits_{0}^{\omega_{\max}}\omega d\omega\left(  1-\frac{\mu}%
{\sqrt{\mu^{2}+\omega^{2}}}\right)
\end{align*}
The last expression is identically coincides with (\ref{117}). We see that
radiation intensity is monotonically grows with the frequency. To evaluate
integral radiation intensity it is the simplest to integrate expression
(\ref{009}):%

\[
I=\frac{b^{2}k_{\max}^{4}}{16\pi mv}=\frac{b^{2}m^{3}(v^{2}-c^{2})^{2}}{\pi
v}.
\]
So, we have the exact agreement of the expressions for the radiation spectrum
of the bose-condensate excitations generated by the fast particle received by
the drag method and the method of the radiated energy flow evaluation. It is
possible to show the ``Fermi's golden rule'' gives the same result. Thus near
the threshold the radiation intensity grows as $(v^{2}-c^{2})^{2}$, but at
high velocity \ --- it is as $v^{3}.$ If in the case of Cherenkov effect the
particle velocity is limited from the top, then for the nonrelativistic
Bose-gas there is no such a limit and with $v\gg c$\ the radiation intensity
can be very big.

We thank RBRF and INTAS for finance support. Grants INTAS-97-0972 and INTAS-97-11066.

\end{document}